\documentclass{desyproc}
\def\ra{\relax\ifmmode{\rightarrow}\else$\rightarrow$\fi}
\def\pim{\relax\ifmmode{\rm \pi^-}\else${\rm \pi^-}$\fi}%
\def\pip{\relax\ifmmode{\rm \pi^+}\else${\rm \pi^+}$\fi}%
\def\pipm{\relax\ifmmode{\rm \pi^{\pm }}\else${\rm \pi^{\pm }}$\fi}%
\def\piz{\relax\ifmmode{\rm \pi^0}\else${\rm \pi^0 }$\fi}%
\def\lb{\relax\ifmmode{\rm \Lambda_b^0}\else${\rm \Lambda_b^0}$\fi}
\def\lc{\relax\ifmmode{\rm \Lambda_c^+}\else${\rm \Lambda_c^+}$\fi}
\def\lcx{\relax\ifmmode{\rm \Lambda_c(2595)^+}\else${\rm \Lambda_c(2595)^+}$\fi}
\def\lcxx{\relax\ifmmode{\rm \Lambda_c(2625)^+}\else${\rm \Lambda_c(2625)^+}$\fi}
\def\scx{\relax\ifmmode{\rm \Sigma_c(2455)}\else${\rm \Sigma_c(2455)}$\fi}%
\def\Pp{\relax\ifmmode{\rm p}\else${\rm p}$\fi}
\def\Pap{\relax\ifmmode{\rm \overline{p}}\else${\rm \overline{p}}$\fi}
\def\Pgmm{\relax\ifmmode{\rm \mu^-}\else${\rm \mu^-}$\fi}%
\def\Pagngm{\relax\ifmmode{\rm \overline{\nu}_{\mu}}\else${\rm \overline{\nu}_{\mu}}$\fi}%
\def\PKm{\relax\ifmmode{\rm K^-}\else${\rm K^-}$\fi}
\def\Pgpp{\relax\ifmmode{\rm \pi^+}\else${\rm \pi^+}$\fi}
\def\pt{\relax\ifmmode{p_{\rm T}}\else$p_{\rm T}$\fi}
\def\rs{\relax\ifmmode{\rm \sqrt{s}}\else${\sqrt{s}}$\fi}
\newcommand{\GeVc}{\ensuremath{{\text{Ge\hspace{-.08em}V\hspace{-0.16em}/\hspace{-0.08em}c}}}}

\newcommand{\GeVcc}{\ensuremath{{\text{Ge\hspace{-.08em}V\hspace{-0.16em}/\hspace{-0.08em}c}^\text{2}}}}
\newcommand{\MeVcc}{\ensuremath{{\text{Me\hspace{-.08em}V\hspace{-0.16em}/\hspace{-0.08em}c}^\text{2}}}}
\begin{document}
\title{First observation and measurement of the resonant structure of the 
$\lb\ra\lc\pim\pip\pim$ decay mode}

\author{{\slshape P.~Azzurri$^1$, P.~Barria$^2$,
M.~A.~Ciocci$^2$, S.~Donati$^3$ and E.~Vataga$^1$,
for the CDF Collaboration.}\\[1ex]
$^1$Scuola Normale Superiore, Piazza dei Cavalieri 7, 56126 Pisa, Italy\\
$^2$Universit\`a di Siena, Dipartimento di Fisica, Via Roma 56, 53100 Siena, Italy \\
$^3$Universit\`a di Pisa, Dipartimento di Fisica, Largo Bruno Pontecorvo 3, 56127 Pisa, Italy}

\contribID{xy}

\confID{800}  
\desyproc{DESY-PROC-2009-xx}
\acronym{LP09} 
\doi  

\maketitle

\begin{abstract}
We present the first observation of the $\lb\ra\lc\pim\pip\pim$ decay
using data from an integrated luminosity of approximately 2.4~fb$^{-1}$
of $\Pp\Pap$ collisions at $\rs$=1.96~TeV, collected with the CDF II 
detector at the Fermilab Tevatron.
We also present the first observation of the resonant decays 
$\lb\ra\scx^0\pip\pim\ra\lc\pim\pip\pim$, 
$\lb\ra\scx^{++}\pim\pim\ra\lc\pim\pip\pim$, 
$\lb\ra\lcx\pim\ra\lc\pim\pip\pim$ and 
$\lb\ra\lcxx\pim\ra\lc\pim\pip\pim$, 
and measure their relative branching ratios.
\end{abstract}

\section{Introduction}
Presented here is the observation of the $\lb\ra\lc\pim\pip\pim$ decay and 
resonant structure in analogy to the 
decay structure observed 
in the  $\lb\ra\lc\pip\pim\Pgmm\Pagngm$ channel~\cite{cdfl}.
All new measurements of the $\lb$ branching ratios 
can be compared to theoretical predictions in the 
heavy quark effective theory (HQEF) approximation~\cite{hqet}. 

This measurement is based on data from an integrated 
luminosity of approximately 2.4~fb$^{-1}$ 
of $\Pp\Pap$ collisions at $\rs$=1.96~TeV, collected with the 
CDF~II detector~\cite{cdf2},
using two-track impact parameter triggers.
Unless stated otherwise, branching fractions, fragmentation functions,
and lifetimes used in the analysis 
are obtained from the Particle Data Group world averages~\cite{pdg}.

\section{Event selection and signal yields}

The event reconstruction and selection has been optimized in order to
maximize the statistical significance of the total number of $\lb$ decays
observed on the data. 
The $\lc$ candidates are reconstructed in the $\lc\ra\Pp\PKm\Pgpp$ channel
requiring a vertex $\chi^2$ probability in excess of 10$^{-4}$,
a transverse decay length in excess of 200~$\mu$m,
$\pt$(p)$>\pt$($\pip$), $\pt(\lc)>4$~$\GeVc$
and the $\lc$ invariant mass in the 2.24-2.33~$\GeVcc$ mass range.

The $\lb$ candidates are reconstructed by further adding to the $\lc$
candidates three pion candidate tracks, with $\eta\phi$-opening 
$\Delta R(3\pi)$ smaller than 1.2.
The $\lb$ candidate is required to have 
a vertex $\chi^2$ probability in excess of 10$^{-4}$,
a transverse decay length in excess of 200~$\mu$m and a 
significance in excess of 16,
an impact parameter smaller than 70~$\mu$m,
and a transverse momentum in excess of 9~$\GeVc$.

\begin{figure}[h]
\begin{minipage}[h]{0.5\textwidth}
\includegraphics[width=1.05\textwidth]{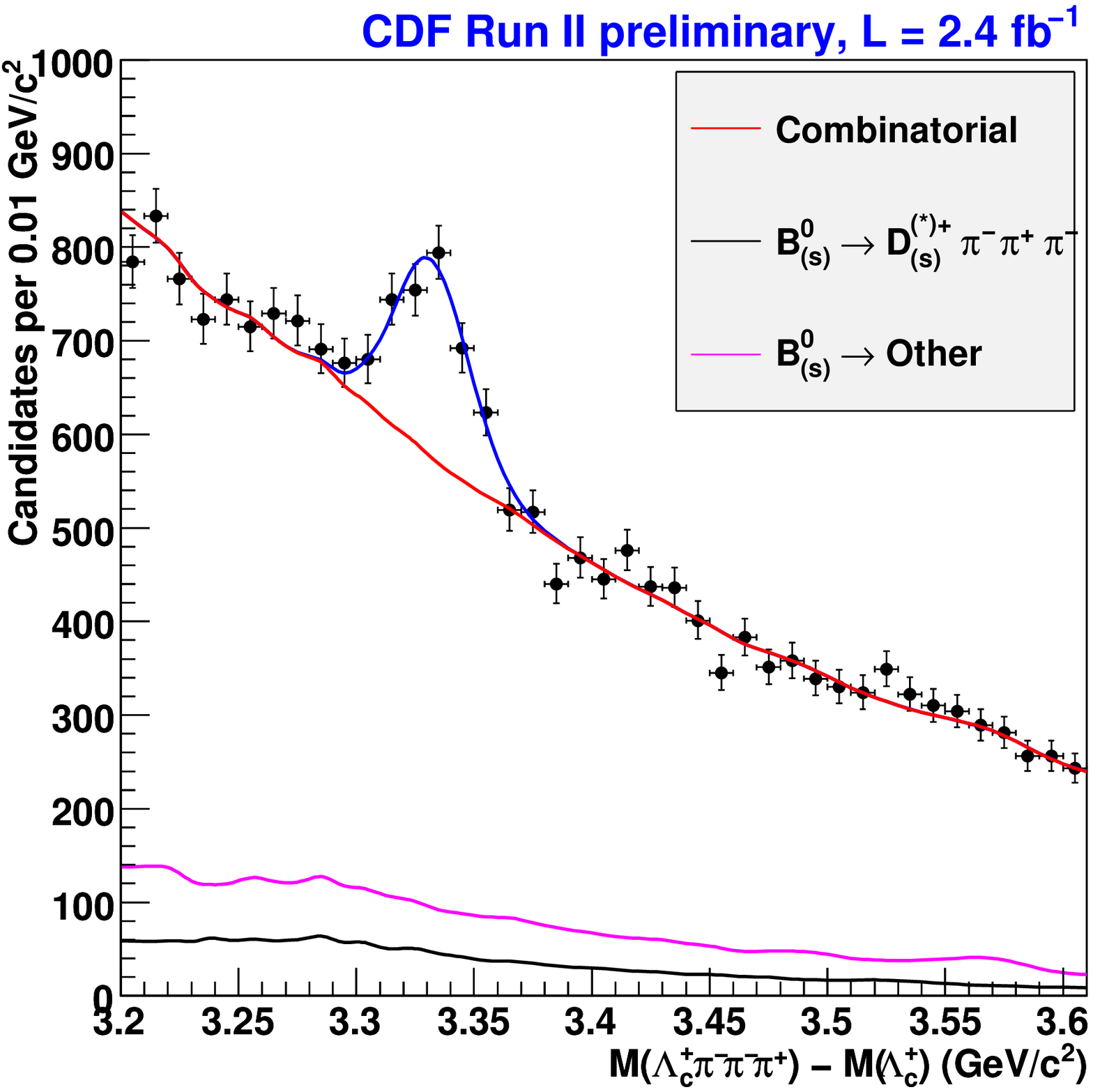}
\caption{The reconstructed invariant mass difference 
$m(\lc\pim\pip\pim)-m(\lc)$, after applying optimized cuts,
showing the total $\lb\ra\lc\pim\pip\pim$ signal yield.}
\label{fig:mlb}
\end{minipage}
\hspace{0.3cm}
\begin{minipage}[h]{0.5\textwidth}
\includegraphics[width=0.95\textwidth]{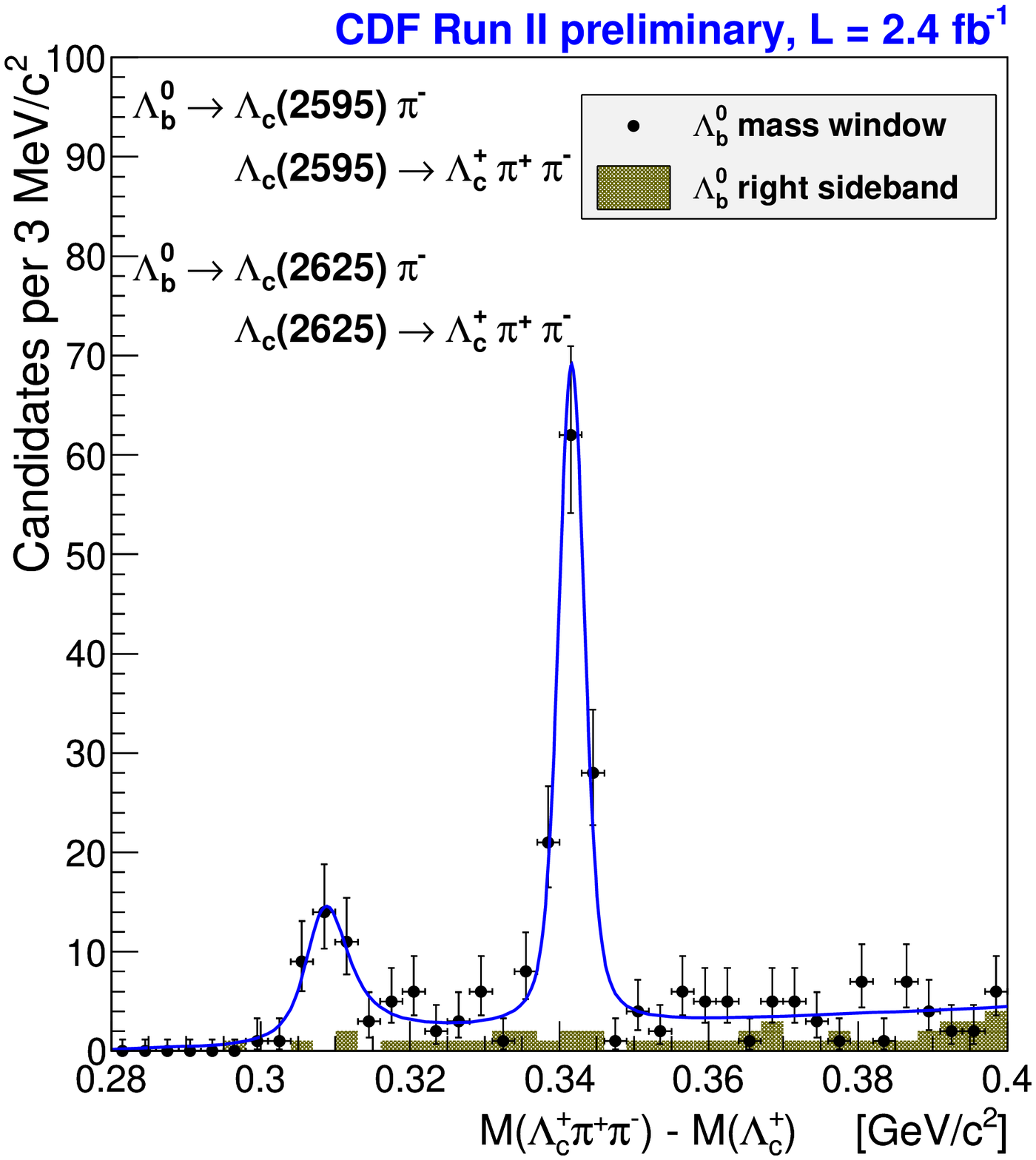}
\caption{The reconstructed invariant mass difference $m(\lc\pim\pip)-m(\lc)$ 
within the $\lb$ mass window, showing the $\lb\ra\lcx\pim\ra\lc\pim\pip\pim$ and 
$\lb\ra\lcxx\pim\ra\lc\pim\pip\pim$ signal yields.}
\label{fig:mlcst}
\end{minipage}
\end{figure}

The resulting distribution of the 
invariant mass difference $m(\lc\pim\pip\pim)-m(\lc)$
with the $\lb\ra\lc\pim\pip\pim$ signal peak, is shown in Figure~\ref{fig:mlb}.
A total signal yield of 848$\pm$93 candidates
is evaluated with an
unbinned likelihood fit using a Gaussian distribution for 
the signal, an exponential distribution for the background,
and Monte~Carlo templates for ${\rm B^0}$ and ${\rm B^0_s}$ backgrounds.
In the following $\lb$ candidates have been selected within 48~$\MeVcc$
of the mass peak.
\vspace{-0.4cm}

The mass difference $\Delta m^{-+}= m(\lc\pim\pip)-m(\lc)$ for selected $\lb$ candidates
is shown in Figure~\ref{fig:mlcst}, with the two peaks from 
$\lcx$ and $\lcxx$ decays. A fit performed with two signal 
peaks and a linear background yields 
46.6$\pm$9.7 $\lb\ra\lcx\pim$ candidates and 
114$\pm$13 $\lb\ra\lcxx\pim$ candidates.

Finally the mass differences $m(\lc\pip)-m(\lc)$ and 
$m(\lc\pip)-m(\lc)$ are shown in Figure~\ref{fig:mlcst},
for selected $\lb$ candidates, after removing $\lcx$ and $\lcxx$ decays
with the $\Delta m^{-+}>360~\MeVcc$ requirement.
Separate fits of the two signal contributions yield 
41.5$\pm$9.3 $\lb\ra\scx^0\pip\pim$ candidates and 
81$\pm$15 $\lb\ra\scx^{++}\pip\pip$ candidates.

\section{Results}
Results are expressed in terms of relative 
branching fractions between the above 
resonant decay modes, correcting for the relative 
channel efficiencies with Monte~Carlo simulations.
Several sources of systematic effects have been
considered, and the dominant uncertainties come 
from the $\lb$ and $\lc$ polarization uncertainty,
and on the unknown fraction of non-resonant 
decays.
 
\begin{figure}[hbt]
\vspace{-0.4cm}
\centerline{
\includegraphics[width=0.5\textwidth]{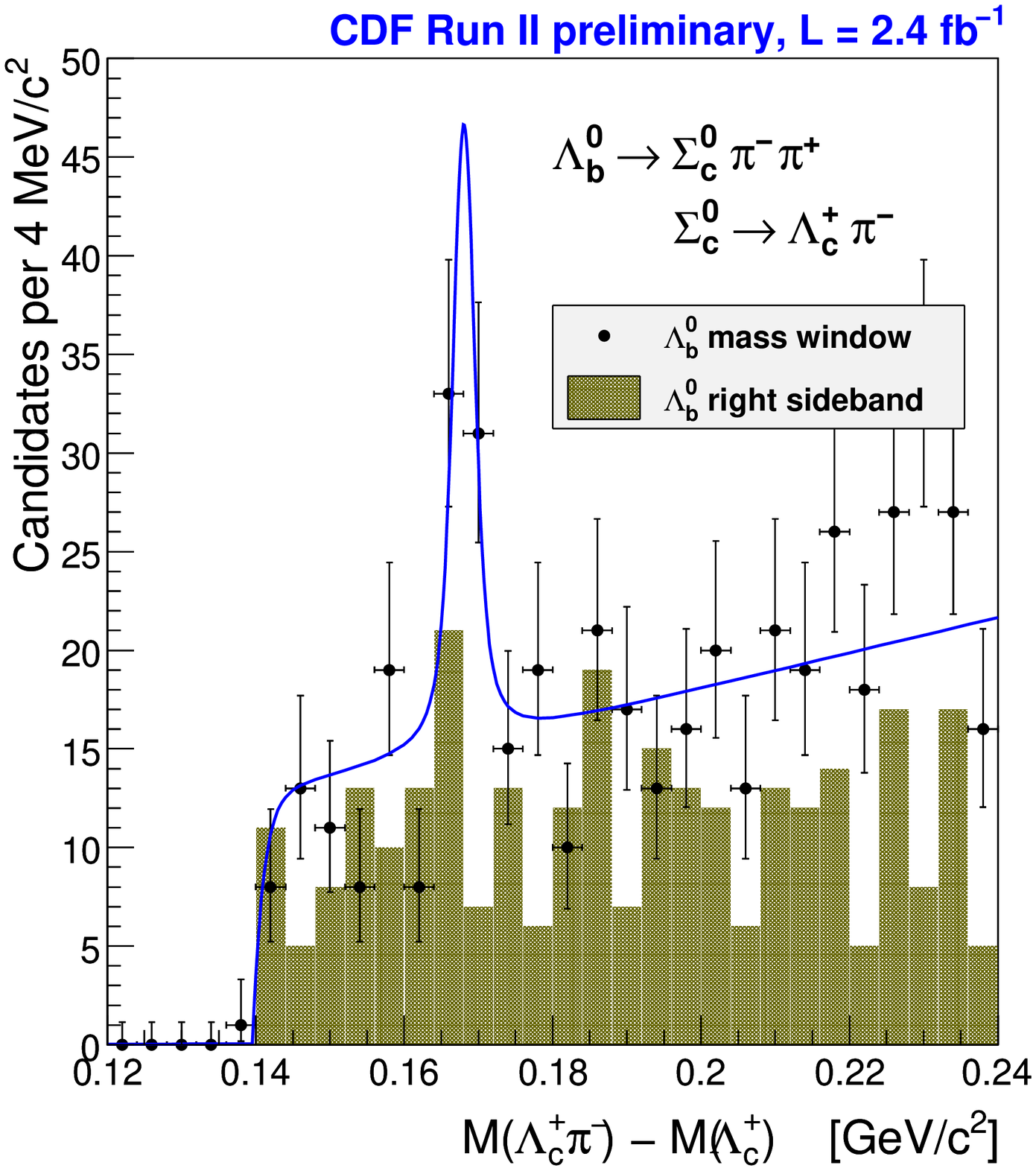}        
\includegraphics[width=0.5\textwidth]{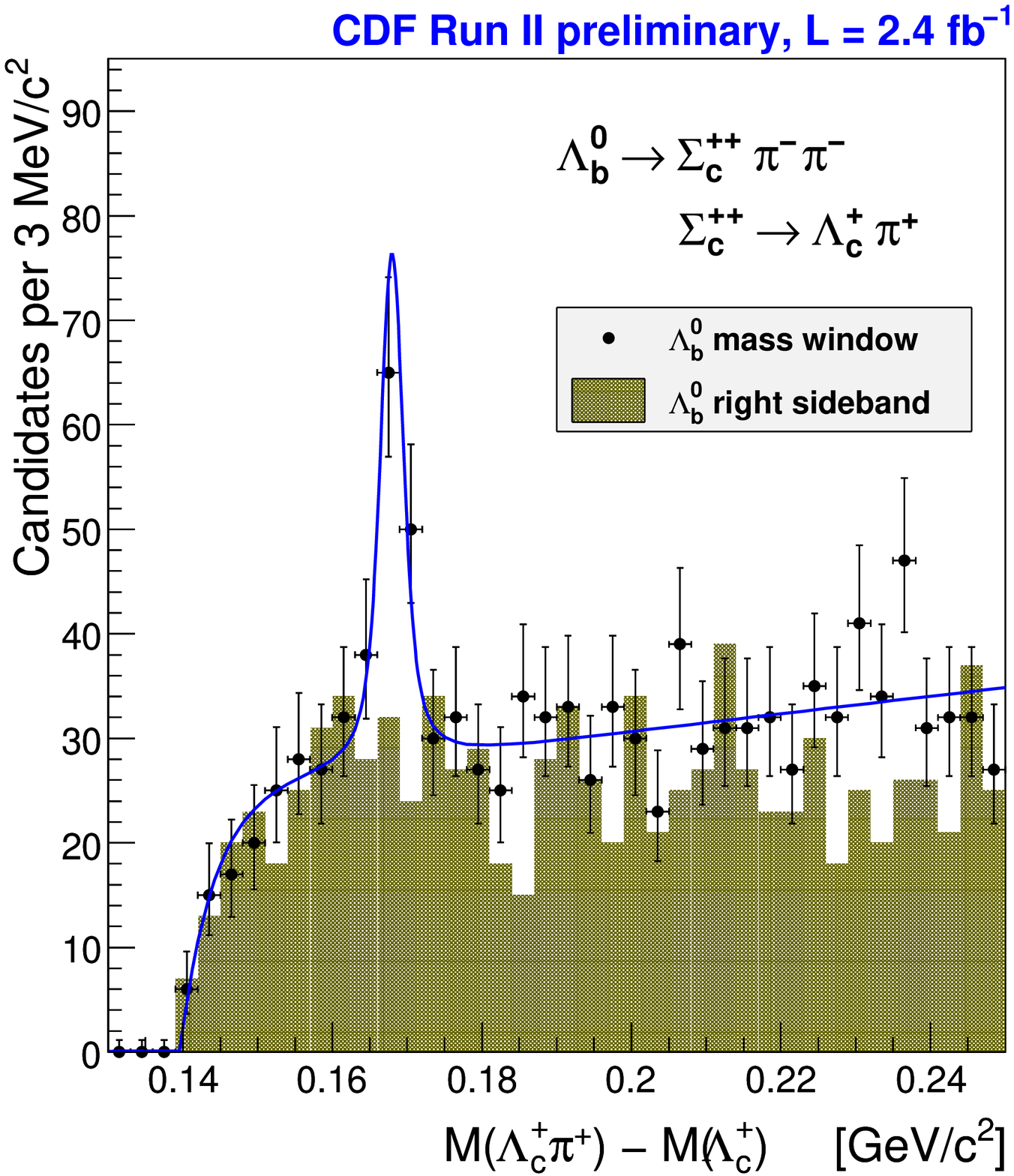}        
}
\caption{ The invariant mass difference $m(\lc\pim)-m(\lc)$ (left)
and $m(\lc\pip)-m(\lc)$ (right) for selected $\lb$ candidates,
after removing events with $\lcx$ and $\lcxx$ decays, and showing 
respectively the presence of $\lb\ra\scx^0\pip\pim$ and  
$\lb\ra\scx^{++}\pim\pim$ signals.}
\label{fig:msc}
\end{figure}

In summary the measured relative branching fractions are the following

\begin{center}

$\frac{\mathcal{B}(\lb\ra \lcx\pim\ra\lc\pim\pip\pim)}
{\mathcal{B}(\lb\ra \lc \pim\pip\pim )} 
= (2.5 \pm 0.6({\rm stat}) \pm 0.5({\rm syst})) \times 10^{-2}$ \\

$\frac{\mathcal{B}(\lb\ra \lcxx\pim\ra\lc\pim\pip\pim)}
{\mathcal{B}(\lb\ra \lc \pim\pip\pim )}
= (6.2 \pm 1.0({\rm stat})^{+1.0}_{-0.9}({\rm syst})) \times 10^{-2}$ \\

$\frac{\mathcal{B}(\lb\ra \scx^{++}\pim \pim\ra\lc\pim\pip\pim)}
{\mathcal{B}(\lb\ra \lc \pim\pip\pim )}
= (5.2 \pm 1.1({\rm stat}) \pm 0.8({\rm syst})) \times 10^{-2}$ \\

$\frac{\mathcal{B}(\lb\ra \scx^{0}\pip \pim\ra\lc\pim\pip\pim)}
{\mathcal{B}(\lb\ra \lc \pim\pip\pim )}
= (8.9 \pm 2.1({\rm stat})^{+1.2}_{-1.0}({\rm syst})) \times 10^{-2}$ \\


$\frac{\mathcal{B}(\lb\ra \lcx\pim
\ra\lc\pim\pip\pim)}
{\mathcal{B}(\lb\ra\lcxx\pim\ra \lc \pim\pip\pim)}
=(40.3 \pm 9.8({\rm stat})^{+2.3}_{- 1.8}({\rm syst}))\cdot 10^{-2}$

$\frac{\mathcal{B}(\lb\ra \scx^{++}\pim\pim
\ra\lc\pim\pip\pim)}
{\mathcal{B}(\lb\ra\scx^0\pip\pim\ra \lc \pim\pip\pim)}
=(58.1 \pm 16.9({\rm stat})^{+6.3}_{-9.1}({\rm syst}))\cdot 10^{-2}$

$\frac{\mathcal{B}(\lb\ra \lcxx\pim
\ra\lc\pim\pip\pim)}
{\mathcal{B}(\lb\ra\scx^{++}\pim\pim\ra \lc \pim\pip\pim)}
=1.20 \pm 0.26({\rm stat})^{+0.05}_{-0.09}({\rm syst})$
\end{center}

where the first error is statistical and the second is from systematic uncertainties.
\vspace{-.2cm}
\begin{footnotesize}

\end{footnotesize}



\begin{thebibliography}{99}

\bibitem{cdfl} T.~Aaltonen {\it et~al.} (CDF Collaboration),  Phys. Rev. {\bf D79} 032001 (2009).
\bibitem{hqet} A.~V.~Manohar and M.~B.~Wise, Cambr. Monogr. Part. Phys. Nucl. Phys. Cosmol. 10, 1 (2000); 
S.~Godfrey and N.~Isgur, Phys Rev. {\bf D32}, 189 (1985); 
N.~Isgur, D.~Scora, B.~Grinstein and M.~B.~Wise Phys Rev. {\bf D39}, 799 (1989); 
A.~K.~Liebovich, I.~W.~Stewart, Phys. Rev. {\bf D57}, 5620 (1998); 
A.~K.~Liebovich, Z.~Ligeti, I.~W.~Stewart, M.~Wise, Phys Lett {\bf B586}, 337 (2004). 
\bibitem{cdf2} D.~Acosta {\it et~al.} (CDF Collaboration),  Phys. Rev. {\bf D71} 032001 (2005).
\bibitem{pdg} C.~Amsler {\it et~al.} (Particle Data Group), Phys, Lett. {\bf B667} 1 (2008).


\end{thebibliography}
\end{document}